\documentclass[11pt]{article}
\usepackage{amsmath, amssymb, amsthm}
\newtheorem{theorem}{Theorem}

\theoremstyle{definition}

\def \FF  {{\cal F}}

\def \MM  {{\cal M}}

\def \PPP {\mathbb{P}}

\def \RRR {\mathbb{R}}
\def \TTT {\mathbb{T}}
\def \ZZZ {\mathbb{Z}}

\title{Convergence of continuous-time quantum walks on the line}
\begin{document}

\author{Alex D. Gottlieb}
\date{}

\maketitle

\begin{abstract}
The position density of a ``particle" performing a continuous-time quantum walk on the integer lattice, viewed on length scales inversely proportional to the time $t$, converges (as $t$ tends to infinity) to a probability distribution that depends on the initial state of the particle.  This convergence behavior has recently been demonstrated for the simplest continuous-time random walk \cite{Konno}.  In this brief report, we use a different technique to establish the same convergence for a very large class of continuous-time quantum walks, and we identify the limit distribution in the  general case.
\end{abstract}

An article recently published in this journal \cite{Konno}
proves that a certain ``continuous-time quantum walk" on the integer lattice $\ZZZ$
has the same kind of convergence behavior as ``coined
quantum walks" on $\ZZZ$ have \cite{K1,K2}.
In this note, the convergence result of \cite{Konno} is generalized to a large class of
continuous-time quantum walks, using the techniques developed in
\cite{GJS,GottliebJansonScudo} for coined quantum walks.
An interesting aspect of the generalized continuous-time quantum walk
is that the limiting measure need not have compact support.

Continuous time quantum walks on graphs were first defined in
\cite{ChildsFarhiGutmann} as follows.  Consider a graph with vertex set $V$.
Let $L_{\gamma}$ denote the infinitesimal generator
of the continuous-time Markov jump process on
$V$, where jumps only occur between adjacent vertices and the
jumping rates are all equal to some $\gamma >0$.  The continuous-time quantum walk of \cite{ChildsFarhiGutmann} amounts to the
unitary dynamics
\begin{equation}
\label{continuousQW1}
    \psi_0 \ \longmapsto \ e^{-itL_{\gamma}}    \psi_0
\end{equation}
on $\ell^2(V)$, the Hilbert space of square-summable complex-valued functions on $V$. Other authors \cite{MooreRussel,GerhardtWatrous} have defined continuous-time quantum walk as the dynamics
\begin{equation}
\label{continuousQW2}
    \psi_0 \ \longmapsto \ e^{-itA}\psi_0\ ,
\end{equation}
using the operator defined by the adjacency matrix $A$ of the graph instead of
$L_{\gamma}$.  When the graph is the integer lattice $\ZZZ$, it hardly matters which way the continuous-time quantum walk is defined, for $L_{\gamma}=(1-2\gamma)I+\gamma A$ and therefore  (\ref{continuousQW1}) and (\ref{continuousQW2}) differ only by a change of time and phase.

In \cite{Konno}, Norio Konno studies the continuous-time quantum walk
on $\ZZZ$ defined as in (\ref{continuousQW2}) but with $-t/2$ instead of $t$.
That is, he studies the dynamics $\psi_0 \mapsto e^{i(t/2)A}\psi_0$,
where $A$ denotes the operator on $\ell^2(\ZZZ)$ whose matrix with respect to the standard orthonormal basis
$\{ e_n \}$ is the adjacency matrix for the integer lattice (the standard basis vector $e_n$ is the member of
$\ell^2(\ZZZ)$ with $e_n(n) = 1$ and $e_n(k) = 0$ for all $k \ne n$).
In this case the operators $e^{i(t/2)A}$ can be expressed exactly in
terms of Bessel functions, and the following convergence behavior
becomes evident \cite{Konno}:   For each $t$, define the probability measures
\[
     P_t(n) \ = \ \big|\langle e_n, e^{i(t/2)A}e_0 \rangle \big|^2
\]
on $\ZZZ$.  Then
\[
    \lim_{t \rightarrow \infty} \sum_{at \le k \le bt} P_t(k) \ =
    \ \int_a^b \frac{dx}{\pi \sqrt{1-x^2}}
\]
for $-1 \le a < b \le 1$.

This result is a special case of a much more general proposition.
The main condition is that the generator $A$ of the quantum walk (\ref{continuousQW2})
be a self-adjoint operator that commutes with translations of $\ell^2(\ZZZ)$.
Any such $A$ has a matrix representation (with respect to the standard basis) of the form
\begin{equation}
\label{generalA}
   \left[
\begin{array}{ccccccc}
\ddots &   \ddots &  \ddots &  \ddots &  \ddots &  \ddots &       \\
\ddots & a_0 & \overline{a_1}  & \overline{a_2}  & \overline{a_3}  & \ddots  & \ddots \\
\ddots & a_1 & a_0  & \overline{a_1}  & \overline{a_2}  & \overline{a_3}  & \ddots \\
\ddots & a_2 & a_1  &  a_0  & \overline{a_1}  & \overline{a_2}   & \ddots \\
\ddots & a_3 & a_2  & a_1 & a_0   & \overline{a_1}  & \ddots \\
\ddots & \ddots  & a_3  & a_2 & a_1 & a_0  & \ddots \\
       & \ddots &  \ddots &  \ddots &  \ddots &  \ddots & \ddots \\
\end{array}
\right]
\end{equation}
with $a_0 = \overline{a_0}$.  We will only consider self-adjoint
operators $A$ that are Fourier transforms of a multiplication
operator, in the following sense. We denote the circle of unit
radius by $\TTT$ and parameterize it by $0 \le \theta < 2\pi$.
$L^2(\TTT, \tfrac{d\theta}{2\pi})$ denotes the Hilbert space of
square-integrable functions $\TTT$. The Fourier transform
\[
     (\FF f)(n) \ = \ \int_{\TTT} f(\theta) e^{-in\theta}
     \tfrac{d\theta}{2\pi}
\]
is a unitary isomorphism from $L^2(\TTT, \tfrac{d\theta}{2\pi})$ to $\ell^2(\ZZZ)$ with inverse
\[
     (\FF^* \psi)(\theta) \ = \ \sum_{n \in \ZZZ} \psi(n) e^{in \theta}\ .
\]
We will assume that there exists a measurable real-valued function
$\widehat{a}(\theta)$ on $\TTT$ such that $A$ satisfies $(\FF^* A
\FF f)(\theta) = \widehat{a}(\theta)f(\theta)$ whenever $f$ and
$\widehat{a}f$ are both in $L^2(\TTT, \tfrac{d\theta}{2\pi})$. We
will further assume that $\widehat{a}(\theta)$ is differentiable
at almost every $\theta \in \TTT$.  These assumptions on the form
of $A$ permit us vastly to generalize the main result of
\cite{Konno} fairly easily, but they are rather technical. Perhaps
a more convenient (but less general) condition is that the entries
of the matrix (\ref{generalA}) satisfy $\sum\limits_{n=1}^\infty
n|a_n|<\infty$, for this implies that $A$ has the desired form
with $\widehat{a}(\theta)=a_0 + \sum\limits_{n=1}^\infty (a_n
e^{in\theta}+ \overline{a_n} e^{-in\theta})$ continuously
differentiable.

The main result of \cite{Konno} is a special case of the following theorem: the case where $\widehat{a}(\theta) = - \cos(\theta)$ and $\psi_0=e_0$.
\begin{theorem}
\label{proveMe}  Suppose that the matrix (\ref{generalA}) represents an operator $A$ on $\ell^2(\ZZZ)$ of the form
$\FF \MM[\widehat{a}]\FF^*$, where $\MM[\widehat{a}]$ denotes the operator on
$L^2(\TTT, \tfrac{d\theta}{2\pi})$ of multiplication by a measurable real-valued function $\widehat{a}(\theta)$.  Suppose that $\widehat{a}'(\theta) = \tfrac{d}{d\theta}\widehat{a}(\theta)$ is defined almost everywhere with respect to Lebesgue measure $d\theta$.   Let $\psi_0$ be any unit vector in $\ell^2(\ZZZ)$ and define
\begin{equation}
\label{P-t}
     P_t(n) \ = \ \left|\langle e_n, e^{-itA} \psi_0 \rangle \right|^2\ ,
\end{equation}
where $e_n$ is the $n^{th}$ standard basis vector in $\ell^2(\ZZZ)$.
Let $\PPP_t[\psi_0]$ be the probability measure
\begin{equation}
\label{KonnoMeasures}
     \PPP_t[\psi_0](dx) \ = \ \sum_{n \in \ZZZ} P_t(n) \delta(x-n/t)
\end{equation}
on $\RRR$, where $\delta(x-n/t)$ denotes the Dirac delta distribution at $n/t$.

Then the
probability measures $\PPP_t[\psi_0]$ converge weakly as $t \longrightarrow \infty$ to the probability measure $\PPP[\psi_0]$
defined on measurable subsets $X \subset \RRR$ by
\begin{equation}
\label{theLimit}
     \PPP[\psi_0](X) \ = \ \frac{1}{2\pi} \int_{ \big\{\theta:\ - \widehat{a}'(\theta) \in X \big\} }
      \big|(\FF^*\psi_0)(\theta)\big|^2 d\theta\ .
\end{equation}
\end{theorem}

\noindent {\bf Proof:} \quad
We will show that the ``characteristic functions" of the probability measures (\ref{KonnoMeasures}) converge to the characteristic function of the probability measure (\ref{theLimit}), for this implies the weak convergence of the  probability measures themselves \cite[Section XIII-1]{Feller}.  The characteristic functions $\Phi_t(\omega)$ of the probability measures $\PPP_t[\psi_0]$ in (\ref{P-t}) are
\begin{eqnarray}
    \Phi_t(\omega)
    & \equiv &
    \int e^{i \omega x}    \PPP_t[\psi_0](dx)
    \ = \ \sum_{n \in \ZZZ} P_t(n) e^{i\omega n/t}
    \ = \ \sum_{n \in \ZZZ}  e^{i\omega n/t} \big|\langle e_n, e^{-itA}\psi_0 \rangle \big|^2  \nonumber \\
    & = &
    \Big\langle e^{-itA} \psi_0,\ E_{\omega/t} e^{-itA} \psi_0 \Big\rangle
    \ = \
    \Big\langle  \psi_0,\ e^{itA} E_{\omega/t} e^{-itA} \psi_0 \Big\rangle
    \label{characteristic} \ ,
\end{eqnarray}
where $E_x$ denotes the multiplication operator
$
     (E_x \psi)(n) = e^{i n x}\psi(n)
$
on $\ell^2(\ZZZ)$.
To show that these characteristic functions converge, we will take Fourier transforms.

Let $\MM[-\widehat{a}']$ denote the multiplication operator
$
    (\MM[-\widehat{a}'] f ) (\theta) =  - \widehat{a}'(\theta) f(\theta)
$
and let
\begin{equation}
\label{H}
H = \FF \MM[-\widehat{a}'] \FF^* \ .
\end{equation}
Note that $H$ may be an {\it unbounded} self-adjoint operator on $\ell^2(\ZZZ)$.
We claim that
\begin{equation}
\label{theClaim}
    \lim_{t \rightarrow \infty} e^{itA}E_{\omega/t} e^{-itA} \psi
    \ = \ e^{i\omega H}\psi
\end{equation}
for all $\psi \in \ell^2(\ZZZ)$.  To prove this claim, first verify that
\begin{eqnarray}
\label{F*AF}
    (\FF^* e^{itA} \FF f)(\theta) & = &  (e^{it\FF^* A \FF} f)(\theta) \ = \ e^{it\widehat{a}(\theta)} f(\theta) \\
\label{F*MF}
    (\FF^* E_{\omega/t} \FF f) (\theta) & = & f(\theta+\omega/t)
\end{eqnarray}
for all $f \in L^2(\TTT, \tfrac{d\theta}{2\pi})$.  Supposing that
$(\FF^*\psi)(\theta)$ is a continuous function on $\TTT$, we calculate that
\begin{eqnarray}
   \lim_{t \rightarrow \infty} (\FF^* e^{itA}E_{\omega/t} e^{-itA} \psi) (\theta)
    & = & \lim_{t \rightarrow \infty} ((\FF^* e^{itA} \FF )( \FF^* E_{\omega/t} \FF )
    ( \FF^*   e^{-itA} \FF)\FF^*\psi) (\theta)
   \nonumber \\
   & = &
   \lim_{t \rightarrow \infty} e^{it\widehat{a}(\theta)}
              e^{-it\widehat{a}(\theta+\omega/t)} (\FF^*\psi)(\theta+\omega/t)
   \nonumber \\
   & = &
   e^{-i\omega \widehat{a}'(\theta)} (\FF^*\psi)(\theta)
   \label{calculation1}
\end{eqnarray}
at almost every $\theta$ by (\ref{F*AF}) and (\ref{F*MF}).  Since $\FF^*\psi$ is bounded, the functions that converge pointwise in (\ref{calculation1}) also converge in $L^2(\TTT, \tfrac{d\theta}{2\pi})$ by Lebesgue's Bounded Convergence Theorem.  The continuity of $\FF$ from $L^2(\TTT, \tfrac{d\theta}{2\pi})$ to $\ell^2(\ZZZ)$ implies that
\[
   \lim_{t \rightarrow \infty} e^{itA}E_{\omega/t} e^{-itA} \psi
  \ = \ \FF e^{i\omega \MM[-\widehat{a}']} \FF^*\psi \ = \ e^{i\omega H} \psi
\]
as claimed in (\ref{theClaim}).  This verifies the claim when $\FF^*\psi$ is a continuous function; the general claim follows by a straightforward density argument.

Applying the claim (\ref{theClaim}) in (\ref{characteristic}) yields
\[
 \lim_{t \rightarrow \infty} \Phi_t(\omega)
   \ = \ \big\langle  \psi_0,\ e^{i\omega H} \psi_0 \big\rangle \ .
\]
Using the unitary isomorphism $\FF$ and the definition (\ref{H}) of $H$ one finds
\begin{eqnarray*}
    \Phi(\omega) & \equiv & \lim_{t \rightarrow \infty} \Phi_t(\omega)
    \ = \ \big\langle  \psi_0,\ e^{i\omega H} \psi_0 \big\rangle_{\ell^2}
    \ = \ \Big\langle  \FF^* \psi_0, \ (\FF^* e^{i\omega H} \FF) \FF^* \psi_0 \Big\rangle_{L^2}
    \\
    & = &
    \Big\langle  \FF^* \psi_0, \ e^{i\omega \FF^* H \FF }  \FF^* \psi_0 \Big\rangle_{L^2}
     \ = \
     \Big\langle  \FF^* \psi_0, \ e^{i\omega \MM[-\widehat{a}'] }  \FF^* \psi_0 \Big\rangle_{L^2}
    \\
    & = &
     \int_{\TTT}
     e^{- i\omega  \widehat{a}'(\theta)} \big| ( \FF^* \psi_0 ) (\theta)\big|^2  \tfrac{d\theta}{2\pi}
\ .
\end{eqnarray*}
This is the characteristic function of the probability measure $\PPP[\psi_0]$ defined in (\ref{theLimit}).
\hfill $\square$


\begin{thebibliography}{X}


\bibitem{Konno} N. Konno.  Continuous-time quantum walk on the line.
To appear in Physical Review E. {\it Manuscript:  quant-ph/0408140}
(2004)


\bibitem{K1} N. Konno.  Quantum random walks in one dimension.
{\it Quantum Information Processing}\ 1: 345-354 (2002)


\bibitem{K2} N. Konno.  A new type of limit theorems for the
one-dimensional quantum random walk.  To appear in Journal of the
Mathematical Society of Japan. {\it Manuscript: quant-ph/0206103}
(2002)


\bibitem{GJS} G. Grimmett, S. Janson, P.F. Scudo.  Weak limits for
quantum random walks. \emph{Physical Review E} 69: 026119 (2004)


\bibitem{GottliebJansonScudo}
A.D. Gottlieb, S. Janson, P.F. Scudo.  Convergence of quantum
walks in $\RRR^d$.  {\it Infinite Dimensional Analysis, Quantum
Probability and Related Topics} 8 (1): 129-140 (2005)


\bibitem{ChildsFarhiGutmann}
A.M. Childs, E. Farhi, S. Gutmann.  An example of the difference
between quantum and classical random walks.  {\it Quantum
Information Processing} 1: 35 - 43 (2002)


\bibitem{MooreRussel}
C. Moore and A. Russell.  Quantum walks on the hypercube. {\it
Proceedings of the Sixth International Workshop on Randomization
and Approximation Techniques in Computer Science} (RANDOM '02)
{\it Manuscript: quant-ph/0104137} (2002)


\bibitem{GerhardtWatrous}  H. Gerhardt and J. Watrous.
Continuous-time quantum walks on the symmetric group. {\it
Proceeding of the Seventh International Workshop on Randomization
and Approximation Techniques in Computer Science}  (RANDOM '03) {\it Manuscript:
quant-ph/0305182} (2003)


\bibitem{Feller}  W. Feller.
 {\it An Introduction to Probability Theory and its Applications II}.
 Second edition, John Wiley \& Sons, New York (1971)







\end{thebibliography}
\end{document}